\documentclass{aa}
\usepackage{graphics}
\begin{document}

\hyphenation{gra-ting pla-ne-ta-ry res-pec-ti-ve-ly mo-dels}

\thesaurus{10.07.2;10.15.1;09.04.1}

\title{New star clusters projected close to the Galactic Centre}

\author{C.M. Dutra \inst{1} \and  E. Bica \inst{1} }

\offprints{C.M. Dutra - dutra@if.ufrgs.br}
 
\institute{Universidade Federal do Rio Grande do Sul, IF, 
CP\,15051, Porto Alegre 91501--970, RS, Brazil}

\date{Received; accepted}

\maketitle

\titlerunning{New clusters near the Galactic Centre}
\authorrunning{C. M. Dutra \& E. Bica}

\begin{abstract}

We carried out a systematic search for new star clusters in a field of 5$^{\circ} \times$5$^{\circ}$ centred  close to the Galactic Nucleus using the infrared JHK$_s$ 2MASS Survey archive.  In addition we searched for embedded clusters in the directions of HII regions and dark clouds for $|\ell|\le4^{\circ}$. As a result we present a list of 58 IR star clusters or candidates. We provide positions, sizes and reddening estimated from 100 $\mu$m dust emission. Their angular distribution together with previously catalogued objects in the region and possible relation with star forming complexes are also discussed.


\end{abstract}

\section{Introduction}

The extinction in regions projected close to the Galactic Centre and Plane made difficult for many years the systematic study of the extended objects therein embedded. However with recent near infrared (NIR) surveys such as the Two Micron All Sky Survey (hereafter, 2MASS; Skrutskie et al. 1997) and the Deep NIR Southern Sky Survey (DENIS; Epchtein et al. 1997) it is  becoming possible to investigate these regions in a spectral domain 10 times less extinguished by dust than the optical. The NIR surveys can provide fundamental data to study the large-scale distribution of galaxies behind the Galactic Plane (Jarret et al. 2000) and the census and distribution of galactic extended objects such as bright, dark and planetary nebulae, globular and open clusters.  Harris \& Racine (1979) estimated that there should be around $\approx$ 160-200 galactic globular clusters.  However so  far there  are 147 known globular  clusters as  indicated in recent compilations (e.g.  Harris 1996 and  updated version  in Web Interface {\it http://physun.physics.
mcmaster.ca/Globular.html}). Thus, new ones could be hidden behind dust clouds in bulge and disk directions. Indeed, Hurt et al. (1999) reported a candidate globular cluster lying only 10$^{\circ}$ away from the Galactic Centre and very close to the plane (b = 0.1$^{\circ}$).  On the other hand, young compact clusters close to the Galactic Nucleus such as  the Arches and Quintuplet clusters (Glass et al. 1990 and Nagata et al. 1995, respectively) as well as embedded clusters in HII regions and dark clouds are interesting objects to be surveyed using NIR images.

In the present study we use the 2MASS survey in the J (1.25$\mu$m), H (1.65$\mu$m) and K$_s$ (2.17$\mu$m) bands to search for potential IR clusters in the central parts of the Galaxy or projected on them. In  Sect. 2 we discuss the process of inspection of 2MASS JHK$_s$ images and present a list of 58 new IR clusters or  candidates. In Sect. 3 we discuss the angular distribution of the sample.  Finally, the concluding remarks are given in Sect. 4.     

\section{IR star clusters or candidates}

 The search was systematically made in the region of 5$^{\circ}$$\times$5$^{\circ}$ centred at 17$^h$51$^m$10$^s$ -28$^{\circ}$16$^{\prime}$10$^{\prime\prime}$ close to the Galactic Centre. In addition we searched for embedded clusters in directions of HII regions and dark clouds for $|\ell|\le4^{\circ}$. In general we considered objects with size and morphology similar to those of the Arches and Quintuplet  which are the closest known clusters to the Galactic Nucleus. We examined a total of 1500 images extracted from the Survey Visualization \& Image Server facility (in the Web Interface {\it http://irsa.ipac.caltech.edu/}). For each available field, we obtained a K$_s$ band image and searched for  objects with dimensions of about 1 arcmin ($\approx$ the Arches' diameter). We extracted new images (JHK$_s$) with $5^{\prime}\times 5^{\prime}$ centred in the coordinates of each  IR cluster candidate from the preliminary list. In this phase we excluded objects affected by artifacts or contaminated by  bright stars on J images. Finally, we obtained a list of 58 objects which are given in Table 1. We determined object positions from K$_s$ images (in FITS format) using {\bf SAOIMAGE 1.27.2 } developed by Doug Mink. We also measured diameters for the objects and their sizes indicate that most of them are suitable only for large ground-based telescopes or Hubble Space Telescope (HST).  Schlegel et al. (1998) built a reddening map from the 100 $\mu$m IRAS dust emission distribution considering temperature effects using 100/240 $\mu$m DIRBE data. Considering our object coordinates, we extracted  reddening values (E(B-V)$_{FIR}$) from  Schlegel et al.'s reddening maps using the software {\bf dust-getval} provided by them. The optical visibility of the IR star clusters or candidates was checked by means of XDSS (Second Generation Digitized Sky Survey) images with $5^{\prime}\times 5^{\prime}$  centred in object position obtained in the Web Interface {\it http://cadcwww.dao.nrc.ca/cadcbin/getdss}.

 Table 1 lists the 58 IR star clusters or candidates, as follows: (1) object identification by a running number along galactic longitude, (2) and (3) galactic coordinates, (4) and (5) equatorial coordinates (J2000 epoch), (6) and (7) the major and minor diameters, (8) optical visibility (yes or no), (9)  E(B-V)$_{FIR}$ reddening values and (10) comments. According to comments in Table 1, we found 20 objects related to or embedded in known emission nebulae (in catalogues L - Lynds 1963, RCW - Rodgers et al. 1960 and Sh - Sharpless 1959), dark nebula (LDN - Lynds 1962) or reflection nebula (Bernes - Bernes 1976). We note that these objects have high E(B-V)$_{FIR}$ values. Since E(B-V)$_{FIR}$ values represent the integrated contribution of the dust along the pathsight in a given direction, it is expected high E(B-V)$_{FIR}$ values in the direction of these star forming complexes close to the Galactic Centre. However, Dutra \& Bica (2000) compared reddening values derived from infrared photometry of embedded clusters in dark clouds with their E(B-V)$_{FIR}$ values and concluded that these reddenings are compatible, except in the Galactic Nuclear region where the temperature in the Central Molecular Zone appears to be underestimated by Schlegel et al.'s temperature maps. High E(B-V)$_{FIR}$ values for objects with traces of optical visibility suggest background dust sources.  
It is interesting to note also that we detect two IR cluster candidates (objects 45 and 46) close to the optical star concentrations NGC\,6432 and NGC\,6465, and two open cluster candidates (objects 18 and 27).    

Figure 1 shows a $3^{\prime}\times 3^{\prime}$ K$_s$ image of the Arches cluster used as reference to search for new clusters close to the Galactic Centre. Figure 2 shows a $3^{\prime}\times 3^{\prime}$ K$_s$ image of the IR star cluster candidate number 11, which is an  embedded cluster candidate in Sh2-21.

\begin{figure} 
\resizebox{\hsize}{!}{\includegraphics{dutraf1.ps}}
\caption[]{$3^{\prime}\times 3^{\prime}$ K$_s$ image of the Arches cluster ($\alpha$ =  17$^h$45$^m$50$^s$ and $\delta$ = -28$^{\circ}$49$^{\prime}$22$^{\prime\prime}$ J2000). }
\label{fig1}
\end{figure}

\begin{figure} 
\resizebox{\hsize}{!}{\includegraphics{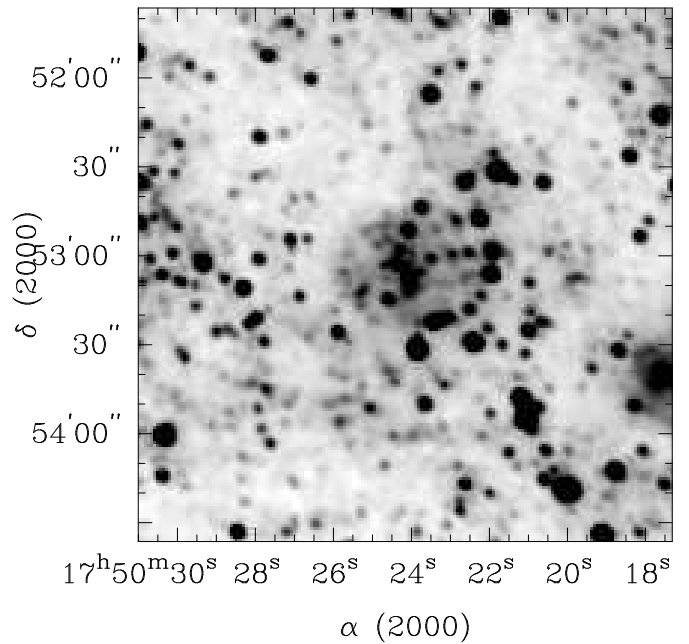}}
\caption[]{$3^{\prime}\times 3^{\prime}$ K$_s$ image of the IR star cluster candidate number 11 ($\alpha$ =  17$^h$50$^m$24$^s$ and $\delta$ = -28$^{\circ}$53$^{\prime}$06$^{\prime\prime}$ J2000).}
\label{fig1}
\end{figure}

\section{Angular distribution}

Figure 3 shows the angular distribution of the IR clusters or candidates compared to that of 58 catalogued open cluster (Alter et al. 1970, Lyng\aa  \,1987, Lauberts 1982) in the 10$^{\circ} \times 10^{\circ}$ region centred on the Galactic Centre. The two known massive compact young clusters Arches and Quintuplet used as references for the search are not indicated, but their galactic coordinates are ($\ell$ = 0.12, {\it b} = 0.01) and ($\ell$ = 0.16, {\it b} =--0.06), respectively. In the systematically surveyed zone (rectangular area) where we detect 58 new IR clusters or candidates there are 24 previously known open clusters (including the Arches and Quintuplet clusters).  We note that there is a deficiency of catalogued open clusters in quadrant Q1, probably caused by nearby dust clouds like those studied by Cambr\'esy (1999).

\begin{figure} 
\resizebox{\hsize}{!}{\includegraphics{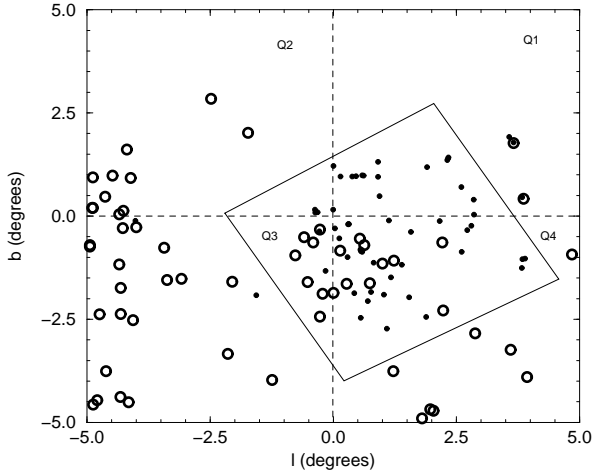}}
\caption[]{Angular distribution of IR clusters and candidates (filled circles) compared to catalogued open clusters (open circles) in the 10$^{\circ} \times 10^{\circ}$ region  centred in the Galactic Centre.  The rectangular area defines the  systematically surveyed region. Galactic Plane and bulge minor axis direction are indicated.}
\label{fig1}
\end{figure}

Figure 4 shows the angular distribution of the IR clusters or candidates compared to 16 known globular clusters in the same region of Figure 3. Only three known globular clusters (Palomar\,6, Terzan\,9 and ESO456SC38) are in the systematically surveyed zone (rectangular area) and we have not seen any additional similar object in the area. This fact could be related to globular cluster destruction due to the tidal effects of the central mass concentration in the Galaxy (Aguilar 1993). Barbuy et al. (1998) studied the spatial distribution of the globular clusters within 5$^{\circ}$ of the Galactic Centre and estimated that there could be 15 missing globular clusters on the opposite side of the Galaxy. They also found evidences of an empty zone inside a radius of about 0.7 kpc,  and that only concentrated clusters would have survived to tidal disruption and disk shocking in central parts of the Bulge.
\begin{figure} 
\resizebox{\hsize}{!}{\includegraphics{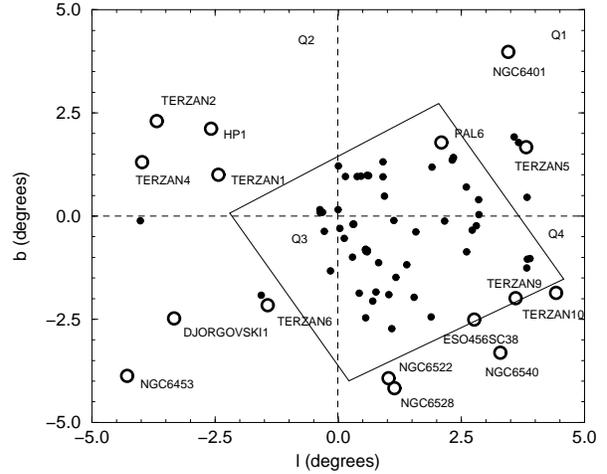}}
\caption[]{Angular distribution of IR clusters and candidates (filled circles) compared to known globular clusters (open circles) in the same region of Figure 3. The surveyed area (rectangular area), Galactic Plane and bulge minor axis direction (dashed lines) are indicated.}
\label{fig2}
\end{figure}

\begin{table*}
\caption[]{New star clusters or candidates}
\begin{scriptsize}
\label{tab1}
\renewcommand{\tabcolsep}{0.9mm}
\begin{tabular}{lrrrrrrrrl}
\hline\hline
Object&$\ell$&{\it b}& RA(2000) & Dec(2000)&Dmax & Dmin&Optical&E(B-V)& Comments\\
&$^{\circ}$&$^{\circ}$&h:m:ss.s~~ &$^{\circ}$:$^{\prime}$~~:$^{\prime\prime}~$&$^{\prime\prime}$&$^{\prime\prime}$~~&Vis.&FIR~~~&\\
\hline
01&  0.03&-0.29&17:46:51.2&-29:03:47& 42& 36&N&26.2&in RCW139                                                  \\
02&  0.11&-0.54&17:48:01.4&-29:06:52& 48& 48&N&10.2&in Sh2-19                                                  \\
03&  0.28&-0.99&17:50:12.4&-29:12:21& 50& 50&Y& 4.0&                                                        \\
04&  0.30&-0.19&17:47:05.7&-28:46:54& 36& 36&Y&57.8&in Sh2-20                                                  \\
05&  0.31&-0.19&17:47:07.0&-28:46:04& 24& 24&Y&55.6&in Sh2-20 deeply embedded, compact few stars                    \\
06&  0.31&-0.20&17:47:09.6&-28:46:26& 45& 36&Y&55.6&in Sh2-20                                                  \\
07&  0.54&-0.81&17:50:06.1&-28:53:13& 48& 48&Y&21.9&at Sh2-21 edge                                             \\
08&  0.55&-0.80&17:50:04.7&-28:52:40& 42& 42&Y&21.9&at Sh2-21 edge                                             \\
09&  0.55&-2.46&17:56:37.5&-29:43:06& 47& 32&N& 1.2&                                                       \\
10&  0.56&-0.85&17:50:17.6&-28:53:40& 36& 36&Y&28.2& in Sh2-21                                     \\
11&  0.58&-0.86&17:50:24.1&-28:53:06& 60& 60&Y&25.4& in Sh2-21                                     \\
12&  0.59&-0.83&17:50:16.4&-28:51:42& 50& 50&Y&21.1& in Sh2-21                                    \\
13&  0.69&-2.05&17:55:20.9&-29:23:26& 43& 37&Y& 1.0&                                                        \\
14&  0.76&-1.84&17:54:38.9&-29:13:36& 47& 30&Y& 1.1&                                                        \\
15&  0.81&-1.12&17:51:56.8&-28:48:56& 40& 33&Y& 1.8&                                                          \\
16&  0.42&-1.86&17:53:58.0&-29:31:42& 39& 39&Y& 1.0&                                                       \\
17&  0.93& 0.48&17:45:57.7&-27:53:16& 45& 28&N&16.9&                                                       \\
18&  0.38& 0.95&17:42:48.4&-28:06:28& 96& 57&Y& 1.8& Open cluster                                                \\
19&  0.90& 1.31&17:42:40.9&-27:28:30& 52& 52&N& 2.0&                                                       \\
20&  0.58& 0.98&17:43:12.4&-27:55:04& 60& 24&N& 2.8&                                                       \\
21&  0.61& 0.98&17:43:17.3&-27:53:45& 51& 22&N& 2.8&                                                       \\
22&  0.46& 0.96&17:42:58.9&-28:02:09& 56& 32&N& 2.6&                                                       \\
23&  0.90& 0.95&17:44:05.2&-27:40:03& 56& 45&Y& 3.8&                                              \\
24&  1.02&-1.90&17:55:30.2&-29:01:39& 56& 56&Y& 1.1&                                                        \\
25&  1.08&-2.73&17:58:54.8&-29:23:40& 43& 43&Y& 1.0&                                                       \\
26&  1.12&-0.10&17:48:41.5&-28:01:42& 40& 35&N&83.1&                                                        \\
27&  1.16&-1.48&17:54:11.6&-28:41:53& 97& 60&N& 1.9& Open cluster                                                  \\
28&  1.39&-1.17&17:53:28.7&-28:20:52& 95& 70&Y& 2.6&                                            \\
29&  0.14& 0.96&17:42:14.1&-28:18:28& 41& 36&N& 2.1&                                                       \\
30&  1.53&-1.97&17:56:55.8&-28:37:21& 43& 30&Y& 1.1&                                                       \\
31&  1.57&-0.38&17:50:49.5&-27:47:07& 47& 47&Y&22.3&                                                        \\
32&  1.88&-2.44&17:59:34.6&-28:33:19& 56& 37&Y& 0.9&                                                        \\
33&  1.90& 1.18&17:45:32.2&-26:41:48& 49& 36&Y& 4.1&                                                       \\
34&  2.16&-0.11&17:51:07.6&-27:08:51& 45& 32&N&37.9&                                                       \\
35&  2.31& 1.36&17:45:48.6&-26:15:03& 48& 36&N& 3.6& related to L9                                         \\
36&  2.33& 1.40&17:45:41.7&-26:12:55& 42& 42&Y& 3.6&includes bright star?, related to L9     \\
37&  2.33& 1.42&17:45:38.0&-26:12:10& 65& 65&Y& 3.5& related  L9     \\
38&  2.59& 0.70&17:48:58.7&-26:21:10& 50& 50&N& 6.0&                                                       \\
39&  2.60&-0.86&17:55:03.3&-27:08:42& 39& 30&N& 7.9&                                                      \\
40&  2.71&-0.34&17:53:15.9&-26:46:52& 36& 21&Y&19.3& few stars                                               \\
41&  2.80&-0.23&17:53:02.8&-26:39:26& 37& 32&N&25.7&                                                       \\
42&  2.84& 0.39&17:50:43.1&-26:17:29& 49& 49&N&18.3&                                                      \\
43&  2.85& 0.03&17:52:07.1&-26:28:18& 67& 64&N&29.2&                                                       \\
44&  3.56& 1.91&17:46:34.9&-24:53:26& 48& 48&Y& 2.7&related to RCW143?                                       \\
45&  3.65& 1.78&17:47:17.5&-24:53:13& 90& 70&Y& 3.3&NW heavily reddened, pair with optical concentration NGC\,6432     \\
46&  3.82& 0.45&17:52:44.5&-25:25:17& 36& 36&Y& 8.7&near optical  concentration NGC\,6465                                      \\
47&  3.82&-1.25&17:59:17.5&-26:17:19& 41& 30&N& 6.2&                                                       \\
48&  3.83&-1.04&17:58:30.0&-26:10:04& 60& 50&N& 8.0&in Reflection  Nebula Bernes 4, in LDN 133                           \\
49&  3.89&-1.03&17:58:34.0&-26:06:55& 54& 42&Y& 8.1&compact few stars, core of LDN\,133, deeply embedded       \\
50&355.98&-0.11&17:36:09.9&-32:24:05&130&130&N&18.3&related to  Sh2-12?                                                \\
51&358.44&-1.91&17:49:29.2&-31:15:51&130&130&Y& 3.5&at edge of Sh2-15, in RCW134                           \\
52&358.78& 0.05&17:42:28.1&-29:56:23& 56& 41&N&37.8&                                                       \\
53&359.56& 0.09&17:44:13.4&-29:15:34& 39& 30&N&45.5&                                                       \\
54&359.62& 0.15&17:44:06.2&-29:10:23& 60& 52&N&25.3&                                                       \\
55&359.63& 0.08&17:44:24.4&-29:12:13& 41& 24&N&39.6&                                                       \\
56&359.71&-0.37&17:46:24.2&-29:22:19& 48& 36&N&27.7&deeply embedded, in Sh2-16       \\
57&359.83&-1.32&17:50:26.5&-29:45:24& 44& 36&N& 1.7&\\
58&359.99& 0.15&17:45:00.1&-28:51:37& 55& 42&Y&61.5&in Sh2-17 \\
\hline
\end{tabular}
\end{scriptsize}
\end{table*}

\section{Concluding remarks}

We provide a list of 58 new IR cluster or candidates detected by means of inspections of 2MASS JHK$_s$ images in the region  5$^{\circ}$$\times$5$^{\circ}$ centred at 17$^h$51$^m$10$^s$ -28$^{\circ}$16$^{\prime}$10$^{\prime\prime}$ close to the Galactic Centre, or in directions of HII regions and dark clouds for $|\ell|\le4^{\circ}$. Most of the objects are structurally  similar to the Arches and Quintuplet clusters. Consequently, they require deep CCD images with large ground-based telescopes or HST to establish their nature. We do not detect any new evident globular cluster in the studied region, which is probably caused by globular cluster destruction due to tidal effects near the Galactic Centre. The angular distribution of the known globular and open clusters in the 10$^{\circ} \times 10^{\circ}$ region centred in the Galactic Centre shows a deficiency of  clusters in quadrant Q1 (0$^{\circ} <$ $\ell$ $< 5^{\circ}$ and 0$^{\circ} <$ {\it b} $< 5^{\circ}$) suggesting a more obscured zone. Infrared surveys such as 2MASS are ideal tools to search for distant new IR open clusters and globular clusters in highly obscured and/or star crowded regions, in particular within 5$^{\circ}$ of the Galactic Centre.

\begin{acknowledgements}

This publication makes use of data products from the Two Micron All Sky Survey, which is a joint project of the University of Massachusetts and the Infrared Processing and Analysis Center/California Institute of Technology, funded by the National Aeronautics and Space Administration and the National Science Foundation.

This publication also use Digitized Sky Survey images for the analysis. The Digitized Sky Survey was produced at the Space Telescope Science Institute
under U.S. Government grant NAG W-2166. The images of these surveys are
based on photographic data obtained using the Oschin Schmidt Telescope on
Palomar Mountain and the UK Schmidt Telescope. The plates were processed into
the present compressed digital form with the permission of these institutions.

We acknowledge support from the Brazilian institution CNPq. 
\end{acknowledgements}

\end{document}